\documentclass[aps,pra,twocolumn,noshowpacs,floatfix]{revtex4-2}

\usepackage{amsfonts}
\usepackage{amssymb}
\usepackage{graphicx}
\usepackage{subfigure}
\usepackage{amsmath}
\usepackage[english]{babel}
\usepackage{color}
\usepackage{epstopdf}
\usepackage{footnote}
\usepackage[utf8]{inputenc}
\usepackage{appendix}
\usepackage{url}

\makeatletter
\renewcommand{\bibsection}{%
	\par
	\onecolumngrid@push
	\vspace{1.5em}%
	\noindent
	\hbox to \textwidth{\hfil\bfseries REFERENCES\hfil}%
	\par
	\vspace{-0.9em}%
	\begingroup
	\baselineskip26\p@
	\bib@device{\textwidth}{245.5\p@}%
	\endgroup
	\nobreak\@nobreaktrue
	\addvspace{4\p@}%
	\par
	\onecolumngrid@pop
}
\makeatother

\begin{document}

\title{Open Quantum Systems Driven by Chirped Pulses: Quantized versus Semiclassical Fields and the Validity of the Rotating-Wave Approximation}

\author{Justin Zhengjie Tan$^{1}$, Frank Großmann$^{2}$, Yiying Yan$^{1,3}$, Maxim Gelin$^{4}$ and Yang Zhao$^{1,}$}
\thanks{Email:~\texttt{YZhao@ntu.edu.sg}}

\affiliation{$^{1}$\mbox{School of Materials Science and Engineering, Nanyang Technological University, Singapore 639798, Singapore}\\
$^{2}$\mbox{Institute for Theoretical Physics, Technische Universität Dresden, 01062 Dresden, Germany} \\
$^{3}$\mbox{Department of Physics, School of Science, Zhejiang University of Science and Technology, Hangzhou 310023, China} \\
$^{4}$\mbox{School of Science, Hangzhou Dianzi University, Hangzhou 310018, China} \\
}

\begin{abstract}
\noindent\textbf{Abstract:}
Population transfer via chirped rapid adiabatic passage is studied using open quantum and semiclassical models, with and without the rotating-wave approximation. A time-dependent variational approach based on the multiple-Davydov D$_2$ trial state is employed to simulate the quantum models with an arbitrary finite mean photon number. We examine the accuracy of both the semiclassical field description and the rotating-wave approximation. Robust population transfer is identified over a wide parameter regime controlled by the laser spectral chirp and is found to be insensitive to the spin--phonon coupling strength, Gaussian pulse area, and energy gap of the two-level system.
\end{abstract}

\maketitle

\begin{center}
\textbf{TOC Graphic}
\vspace{0.2cm}
\includegraphics[width=3.25in,height=1.75in,keepaspectratio]{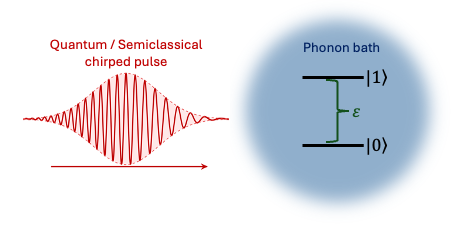}
\end{center}


Electromagnetic driving provides a widely used experimental handle to coherently shape quantum dynamics and has become foundational in quantum optics \cite{kimble,Ritsch}, quantum control \cite{CDT,Wollenhaupt,Brif,Keefer,Hiroyuki}, and atomic and molecular physics \cite{Metcalf,Bergmann,Shapiro,Tannor,Shore}. In the large mean photon number and weak system-photon coupling limit, both the semiclassical and quantum descriptions of the light field are able to accurately capture the system dynamics \cite{YYY_ZG}. However, in the low mean photon number regime, field quantization and its associated fluctuations are imperative for quantitative accuracy. Mapping the parameter regimes where the two radiation descriptions converge or diverge is therefore key both for conceptual clarity and for robust experimental design. 

Introduced in nuclear magnetic resonance \cite{Bloch,Bloch_Hansen}, the rapid adiabatic passage (RAP) offers robust population transfer and is less sensitive to experimental imperfections than a $\pi$-pulse protocol \cite{Melinger,Malinovsky}. In RAP, the carrier frequency of the chirped radiation pulse is a slowly varying function of time, which is swept through resonance. Recently, theoretical efforts \cite{Debnath,Werther_Grossmann} have been made to replicate the experimental findings \cite{Simon_Bel} in RAP for semiconductor quantum dots to a large degree of success. Key findings show that for odd integers of $\pi$ pulse area, combined with small chirp parameter, are ideal for optimal population transfer. Also, strong system-phonon coupling strength assists in the stabilization of transition probability. These results were obtained within the assumption of the rotating-wave approximation (RWA) and with a semiclassical treatment of the field. Studies of RAP dynamics without RWA and a quantized description of a field in the presence of dissipation are lacking in the current literature landscape.

Motivated by the ubiquity of loss and decoherence in experimental platforms, we thus study RAP in the open Rabi model and the open Jaynes--Cummings (JC) model, which are two paradigmatic models \cite{JCmodel} that provide a more realistic description than idealized closed system treatments \cite{AllenEberly,Vitanov,RangelovSCRAP3,TorosovCAP,MiaoARPForce,AmniatTalabCavitySCRAP,WeiSCRAP_SCQubits,Nalbach,ChenWeiCircuitQEDSCRAP}. Under the RWA, where the quickly oscillating counter-rotating terms are omitted, the quantum JC model can be derived from the quantum Rabi model and a similar limit also exists for their semiclassical counterparts. 
In this study, we will compare the semiclassical and the quantum treatments of the light field, with and without RWA. In this way, four relevant models (quantum and semiclassical Rabi models, along with quantum and semiclassical JC models) are considered. Our numerics address whether the robustness of population transfer previously reported in a semiclassical RWA treatment \cite{Werther_Grossmann} persists in a fully quantum, open-system description, thereby establishing its potential experimental viability and utility for dissipative state engineering \cite{Lin}.

With a semiclassical treatment of a chirped laser pulse interacting with a spin-1/2 system in the presence of an external bath, the Hamiltonian can be described as follows
\begin{equation}
\begin{aligned}
&H_{\mathrm{SC}} = \frac{\varepsilon}{2}\sigma_z
+ \sum_{k=1}^{N}\omega_k\, b_k^\dagger b_k
+ \sigma_z \sum_{k=1}^{N}\frac{\lambda_k}{2}\left(b_k+b_k^\dagger\right) \\
&+ \frac{\eta(t)}{2}\Big[ \sigma_+ e^{-i\Phi(t)} + \sigma_- e^{i\Phi(t)}
+ \zeta\big(\sigma_+ e^{i\Phi(t)} + \sigma_- e^{-i\Phi(t)}\big)\Big]
\end{aligned}
\end{equation}

where

\begin{eqnarray}
    \Phi(t)&=&\int^t_{t_0} \omega(\tau)d\tau, \\
    \omega(t) &=& \omega_0 + 2\beta t.
\end{eqnarray}

The chirp is encoded through the accumulated phase is denoted as $\Phi(t)$ and the instantaneous laser frequency is given by $\omega(t) = \omega_0 + 2\beta t$, where $\omega_0$ is the carrier frequency and $\beta$ is the linear temporal chirp \cite{Chatel_Degert,Darsheshdar}. $\sigma_{x,z}$ are the Pauli matrices and $\sigma_{+,-}$ are the spin raising and lowering operators for a two-level system. The phononic bath mode of the $k$-th frequency is represented by the creation operator $b_k^\dagger$. $\varepsilon$ represents the energy gap of the spin. The laser and the energy gap of the spin are assumed to be resonant at $t=0$, i.e., $\varepsilon=\omega_0$. $\eta(t)$ denotes the time-dependent Gaussian envelope of the electric field
\begin{eqnarray}\label{Eq_Gauss}
    \eta(t) = \frac{\Theta}{\mu\sqrt{\pi}\tau_p} \exp \Big( -\frac{t^2}{\tau_p^2} \Big),
\end{eqnarray}
where $\tau_p$ is the pulse duration, and the peak amplitude $\frac{\Theta}{\mu\sqrt{\pi}\tau_p}$ is scaled with the dipole matrix element $\mu$ and is characterized by the pulse area $\Theta$. The frequency bandwidth $\Gamma$ and the spectral chirp $\phi^{\prime\prime}$ are related to the pulse duration and the spectral temporal chirp via \cite{Malinovsky}
\begin{eqnarray}
    \tau_p^2 = \frac{1}{\Gamma^2}\big[ 1+(2\phi^{\prime\prime})^2\Gamma^4 \big],
\end{eqnarray}
\begin{eqnarray}
    \beta = \frac{2\phi^{\prime\prime}\Gamma^4}{1+(2\phi^{\prime\prime})^2\Gamma^4}.
\end{eqnarray}

The mode-dependent $\lambda_k$ can be extracted from the continuous spectral density function
\begin{eqnarray}\label{eq_SD}
    J(\omega) = A\omega^3\exp (-\frac{\omega^2}{\omega_c^2}) \approx \sum_{k=1}^N \lambda_k^2\delta(\omega-\omega_k),
\end{eqnarray}
where $A$ specifies the coupling strength and $\omega_c$ represents the cutoff frequency. The spectral density is discretized in the frequency domain and the number of bath degrees of freedom is treated explicitly on the wavefunction level together with the system's dynamics.

Finally, for $H_{\mathrm{SC}}(\zeta=0)$, the counter-rotating terms are discarded and the Hamiltonian becomes the open semiclassical JC model ($H_\text{SCJC}$) whereas $H_{\mathrm{SC}}(\zeta=1)$ recovers the open semiclassical Rabi model ($H_\text{SCRM}$).

The effective Hamiltonian for the full quantum treatment of a chirped laser pulse of the same composite system shall be expressed as 
\begin{eqnarray}
    H_\text{Q} &=& \frac{\varepsilon}{2}\sigma_z + \sum_{k=1}^N \omega_kb_k^\dagger b_k + \sigma_z\sum_{k=1}^N \frac{\lambda_k}{2} \Big( b_k+b_k^\dagger \Big) \nonumber\\
    + \omega(t)a^\dagger a &+& \frac{\eta(t)}{2|\alpha|} \Bigg[ a\sigma_+ + a^\dagger\sigma_- ~+~ \zeta\Big( a^\dagger\sigma_+ + a\sigma_- \Big) \Bigg]. \nonumber\\
\end{eqnarray}
For $H_\mathrm{Q}(\zeta=0)$, the counter-rotating terms are discarded and the Hamiltonian becomes the open quantum JC model ($H_\text{QJC}$) whereas $H_\mathrm{Q}(\zeta=1)$ recovers the open quantum Rabi model ($H_\text{QRM}$). The single photonic creation operator is expressed as $a^\dagger$. The reason behind representing the quantum chirped pulse through a time-dependent spin-photon coupling is driven by a mathematical perspective \cite{Irish_Armour} and has a precursor in the JC model with atomic motion introduced by Schlicher \cite{Schlicher}. After applying two sequential unitary transformations to the quantized driving term, the semiclassical driving term can be recovered with an additional quantum fluctuation term which is shown in the Supporting Information. The quantum field model used here should be interpreted as an effective single-mode representation of a chirped pulse. The Gaussian envelope is encoded through the time-dependent coupling \(\eta(t)\). This construction is designed to reproduce the semiclassical chirped drive in the large-\(|\alpha|^2\) limit, rather than to represent a full multimode propagating laser pulse.

The time evolution of the composite system is governed by the time-dependent Schrödinger equation,
\begin{eqnarray}
    i\frac{d}{dt}|\Psi (t)\rangle = H|\Psi (t)\rangle,
\end{eqnarray}
where $|\Psi (t)\rangle$ is the state of the total system. In this work, we consider a factorised initial state of the total system 
\begin{eqnarray}
    |\Psi (t\to-\infty)\rangle = |\psi(t\to-\infty)\rangle \otimes |\alpha\rangle,
\end{eqnarray}
where $|\psi(t\to-\infty)\rangle$ is an initial state of the spin-bath system and $|\alpha\rangle$ is a coherent state that is expressed as \cite{Glauber}
\begin{eqnarray}
    |\alpha\rangle \equiv \exp(\alpha a^\dagger - \text{H.c.})|\textbf{0}\rangle \equiv D(\alpha)|\textbf{0}\rangle,
\end{eqnarray}
where $|\textbf{0}\rangle$ is the photon mode vacuum state, $D(\alpha)$ is a displacement operator, and $\alpha\equiv |\alpha|e^{-i\varphi}$ is a complex number with modulus $|\alpha|$ and phase $\varphi$. $\varphi=0$ is fixed throughout the study. 

To achieve a manageable numerical simulation, we convert the time-evolution problem with the initial coherent state of a large number of bosons into a new time evolution problem with an initial vacuum state. This can be achieved with two sequential unitary transformations. First, we transform the time-dependent Schrödinger equation into the interaction picture governed by the total bosonic field, followed by a displacement transformation as shown in \cite{YYY_ZG}. Next, a time-dependent variational approach is employed to compute the dynamics. The variational approach is based on the Dirac-Frenkel time-dependent variational principle and the multiple Davydov D$_2$ ansatz (mD$_2$) \cite{Zhao_Yang,Zhao_Yang2}. The latter is expressed as \cite{Frenkel}
\begin{eqnarray}
    |{\rm D}_2^M(t)\rangle = \sum^M_{n=1}\sum^{N_s-1}_{j=0} A_{nj}(t)|j\rangle|f_n(t)\rangle,
\end{eqnarray}
where $M$ is the number of coherent states, $A_{nj}(t)$ are time-dependent variational parameters, $\{ |j\rangle|j=0,1,2,...,N_s-1\}$ represents a set of basis bases for the quantum system, and 
\begin{eqnarray}
    |f_n(t)\rangle=\exp \bigg( \sum^{N+1}_{k=1} f_{nk}(t)b^\dagger_k - \text{H.c.} \bigg)|\textbf{0}\rangle
\end{eqnarray}
are the bosonic states with $f_{nk}(t)$ being time-dependent variational parameters. Employing the Dirac-Frenkel time-dependent variational principle, the equations of motion for the variational parameters are shown in the Supporting Information.

Note that all parameters are made dimensionless with respect to the phonon cutoff frequency $\omega_c$, i.e., $\varepsilon\in[10^0,10^3]\omega_c$ and $\phi^{\prime\prime}\in[0,-40]\omega_c^2$.

\begin{figure*}
    \centering
	\includegraphics[width=\textwidth, height=0.3\textheight]{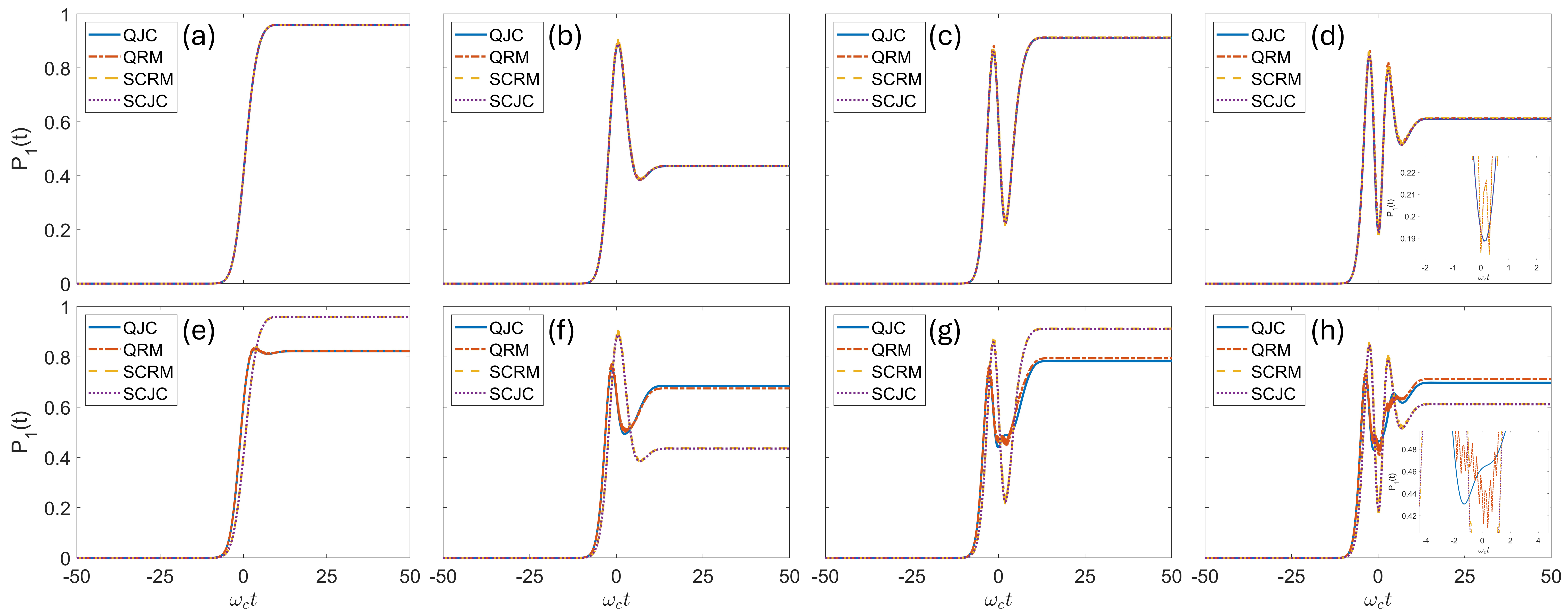} \\
    \caption{Time evolution of population transfer $P_1(t)$ of the quantum JC model (blue), quantum Rabi model (orange), semiclassical Rabi model (yellow) and semiclassical JC model (purple) with different initial mean photon number $|\alpha|^2$ for various pulse area: (a) $|\alpha|^2=10^4$, $\Theta=\pi$; (b) $|\alpha|^2=10^4$, $\Theta=2\pi$; (c) $|\alpha|^2=10^4$, $\Theta=3\pi$; (d) $|\alpha|^2=10^4$, $\Theta=4\pi$; (e) $|\alpha|^2=1$, $\Theta=\pi$; (f) $|\alpha|^2=1$, $\Theta=2\pi$; (g) $|\alpha|^2=1$, $\Theta=3\pi$; and (h) $|\alpha|^2=1$, $\Theta=4\pi$. The spin-phonon coupling $A$ is set to be $0.22~\omega_c^{-2}$, the spectral chirp $\phi^{\prime\prime}=-7~\omega_c^2$ and the energy gap at resonant excitation ($\varepsilon=\omega_0$) is set at $10~\omega_c$. }
    \label{figure1}
\end{figure*}

Let us first look at the time evolution of the population transfer $P_1(t)$ across all four models, with various pulse areas, in the high and low $|\alpha|^2$ regime, as illustrated in Fig. \ref{figure1}. The calculations are performed for the representative values of spin-phonon coupling and chirp, while a detailed analysis of the impact of these quantities on the population transfer is given below. The combination of high $|\alpha|^2$ and small pulse area leads to good agreement across all four models at all times, as shown in Fig. \ref{figure1}(a), with all four curves indistinguishable. This can be attributed to the fact that the quantum and semiclassical dynamics are consistent in the large mean photon number and weak coupling limit \cite{YYY_ZG}. The RWA is accurate in this parameter space as well. With high $|\alpha|^2$ and increasing pulse area, the Rabi models start to diverge from the JC models around $\omega_ct=0$ (which is around the maximum amplitude of the coupling strength $\eta$), as depicted in Fig. \ref{figure1}(b)-(d). This implies that the RWA is breaking down at intermediate times even though the final-state population $P_f$, where $P_f:= P_1(t\to\infty)$, is the same. 

For low $|\alpha|^2$, the semiclassical models are unable to simulate the quantum mechanical effects of the light field, which explains the difference of $P_1(t)$ between the quantum and semiclassical models. The drastic disagreement of semiclassical and quantum models is depicted across all pulse areas. In the effective quantum Hamiltonian in Eq. (6), a lower $|\alpha|^2$ magnitude might seem to increase spin-photon coupling strength. However, as shown explicitly in the Supporting Information, after two sequential unitary transformations, the quantum spin-photon coupling term can be split into two terms. The first term recovers the original semiclassical driving term $H^{SC}_\text{drv}$ and the second term is a quantum fluctuation term $H_\text{fluc}$, where
\begin{align}
H_{\mathrm{fluc}}(t)
&=
\frac{\eta(t)}{2|\alpha|}
\Big[
ae^{-i\Phi(t)}\sigma_{+}
+
a^{\dagger}e^{i\Phi(t)}\sigma_{-}
\nonumber\\
&\hspace{1.4cm}
+
\zeta
\big(
a^{\dagger}e^{i\Phi(t)}\sigma_{+}
+
ae^{-i\Phi(t)}\sigma_{-}
\big)
\Big].
\end{align}
The quantum fluctuation term scales inversely with $|\alpha|$, where a low $|\alpha|^2$ leads to a stronger quantum fluctuation effect. Conversely, a high $|\alpha|^2$ will cause the quantum fluctuation term to be trivial and recover the semiclassical driving term only. Furthermore, in \cite{YYY_ZG}, it is found quantitatively in a closed JC model that the second-order quantum corrections can contribute significantly to the divergence of the quantum and semiclassical dynamics even when the photon-spin coupling is small. It is also suggested that higher-order quantum corrections must be included to account for the quantum-classical difference as $t\to\infty$.

The beyond-RWA treatment of Browne and Keitel \cite{Browne} can be adapted to the present RAP setting as an instantaneous perturbative picture. Their central idea is that when the Rabi frequency becomes comparable to the drive frequency, counter-rotating terms can no longer be neglected because they mix neighboring dressed states and shift the dressed energies. In our chirped model, this translates to replacing the monochromatic laser by Eq. 3 and $\Omega_{Rabi}(t)\sim\eta(t)$, so that RWA breakdown is expected when $\eta(t)/\omega(t)$ is no longer small, especially near the pulse center where $\eta(t)$ reaches its maximum. This explains why the QRM and QJC results begin to diverge around $t=0$ and at larger pulse areas. In the QJC model, only excitation-number conserving processes are allowed such as $|0,n+1\rangle \leftrightarrow |1,n\rangle$ (spin excitation with photon annihilation or spin relaxation with photon creation) whereas QRM allows excitation-number non-conserving processes such as $|0,n\rangle \rightarrow |1,n+1\rangle$ (simultaneous spin excitation and photon creation) and $|1,n\rangle \rightarrow |0,n-1\rangle$ (simultaneous spin relaxation and photon annihilation). These additional counter-rotating terms in the QRM ($a^\dagger \sigma_+ + a \sigma_-$), that are absent in the QJC model, renormalize the instantaneous dressed-state structure and generate excitation-number non-conserving pathways that lead to the divergence in results. Nevertheless, because the present problem is explicitly time dependent and includes both phonon dissipation and low-photon-number fluctuations, the perturbative dressed-state picture serves only as an interpretation of the onset of non-RWA effects, whereas the full dynamics must still be obtained from the time-dependent variational treatment.

\begin{figure}
    \centering
    \includegraphics[width=\linewidth]{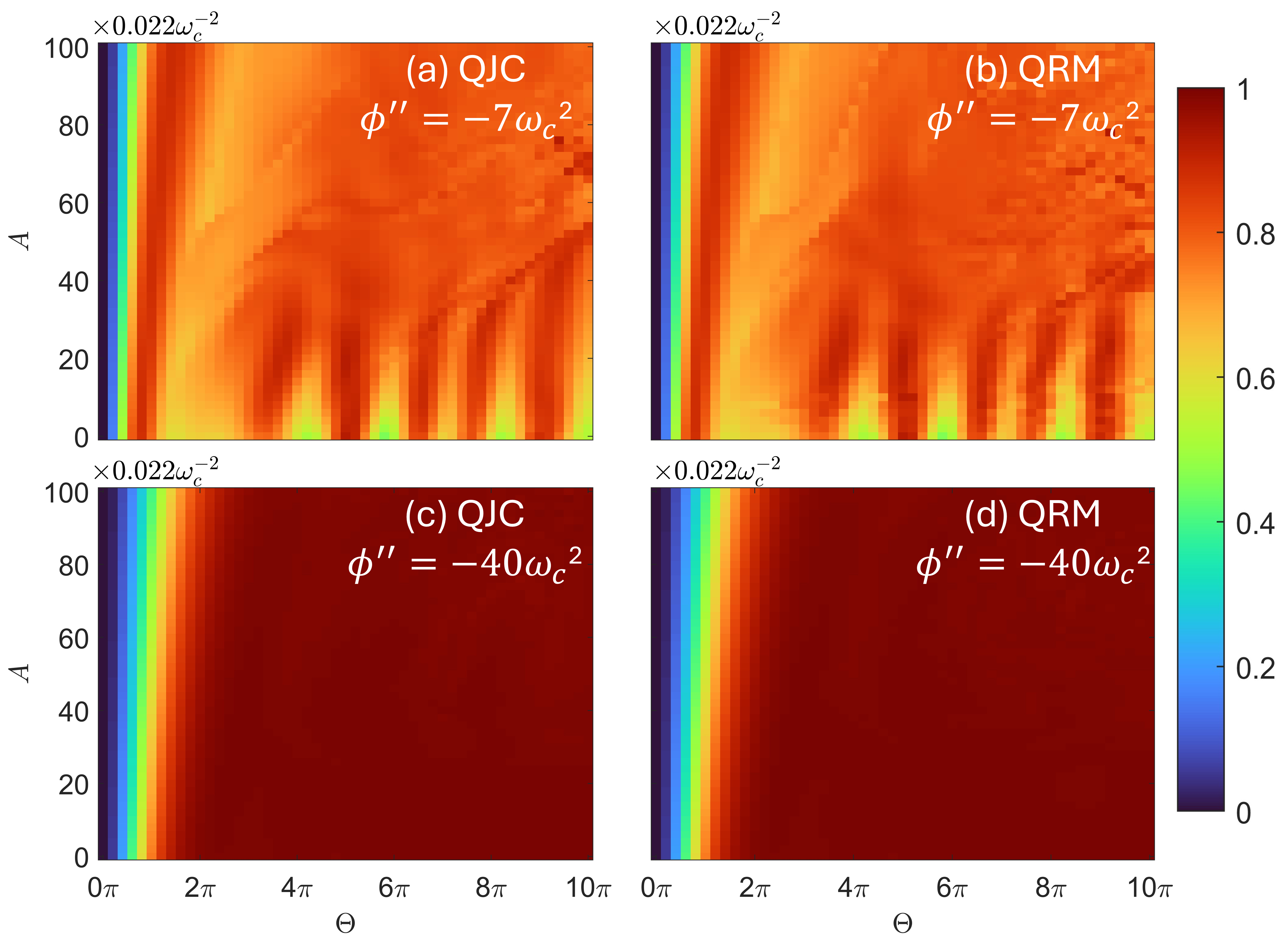}
    \caption{Two-dimensional heatmaps of the final transition probability $P_f$ as a function of pulse area $\Theta$ and spin-phonon coupling $A$. The top row depicts the heatmaps for the spectral chirp $\phi^{\prime\prime}=-7~\omega_c^{2}$ for (a) QJC and (b) QRM. The bottom row shows the heatmaps for the spectral chirp $\phi^{\prime\prime}=-40~\omega_c^{2}$ for (c) QJC and (d) QRM. The following parameters are set at: $|\alpha|^2=1$ and $\varepsilon=\omega_0=10~\omega_c$.}
    \label{figure2}
\end{figure}

In Fig. \ref{figure2}, the final-state population $P_f$ is plotted as a function of pulse area and spin-phonon coupling $A$. Additionally, the effect of various spectral chirps $\phi^{\prime\prime}$ is studied. In Fig. \ref{figure2}(a) and (b), when $\phi^{\prime\prime}=-7~\omega_c^2$ in the low $A$ regime, odd-integer multiples of $\pi$ in the pulse area lead to near complete population transfer i.e., $P_f\approx 1$. Previous studies done in this parameter regime have reported the same phenomenon \cite{Werther_Grossmann,Debnath,Simon_Bel}. Both the QRM and QJC models are in good agreement when $\Theta\leq5\pi$. As $\Theta>5\pi$, the absence of counter-rotating terms in the QJC model due to the RWA starts to affect $P_f$, as is evident when $P_f$ is compared with that obtained from the QRM. When $\phi^{\prime\prime}=-40~\omega_c^2$, both the QRM and QJC models obtain nearly identical $P_f$. However, the high $P_f$ phenomenon from the odd-integer multiples of $\pi$ in the pulse area vanishes completely.

As $|\phi^{\prime\prime}|$ increases from $|-7~\omega_c^2|$ to $|-40~\omega_c^2|$, the linear temporal chirp $\beta$ decreases from $|-0.0171~\omega_c^{2}|$ to $|-0.0114~\omega_c^{2}|$. A popular tool to estimate the transition probability between two crossing adiabatic states is the Landau-Zener formula \cite{Landau,Zener,Vitanov}
\begin{eqnarray}
    P_f=1-\exp\bigg[-\frac{\pi\Omega^2_{t=0}}{2|\dot{\Delta}_{t=0}|}  \bigg], \nonumber\\
\end{eqnarray}
where $|\dot{\Delta}_{t=0}|=|2\beta|$ is the rate of change in the detuning and $\Omega^2_{t=0}$ is the Rabi frequency at the crossing time $t=0$. With a smaller magnitude of $\beta$, the instantaneous frequency sweep is slower and the adiabatic parameter $\Lambda_{LZ}=\Omega^2_{t=0}/|\dot{\Delta}_{t=0}|$ increases which allows the system to follow the adiabatic basis better. This leads to a higher transition probability which agrees with the mD$_2$ results in Fig. \ref{figure2}. Even in the presence of a diagonally coupled phonon mode, the analytical formula of the Landau-Zener transition probability remains unaffected \cite{HZK}. When the spectral chirp $\phi^{\prime\prime}=-40~\omega_c^2$, $P_f$ is largely independent of the pulse area as the pulse area increases beyond $3\pi$. Robust maximum population transfer can be achieved ($P_f \approx 1$) irrespective of the coupling to the environment.

\begin{figure}
    \centering
    \includegraphics[width=\linewidth]{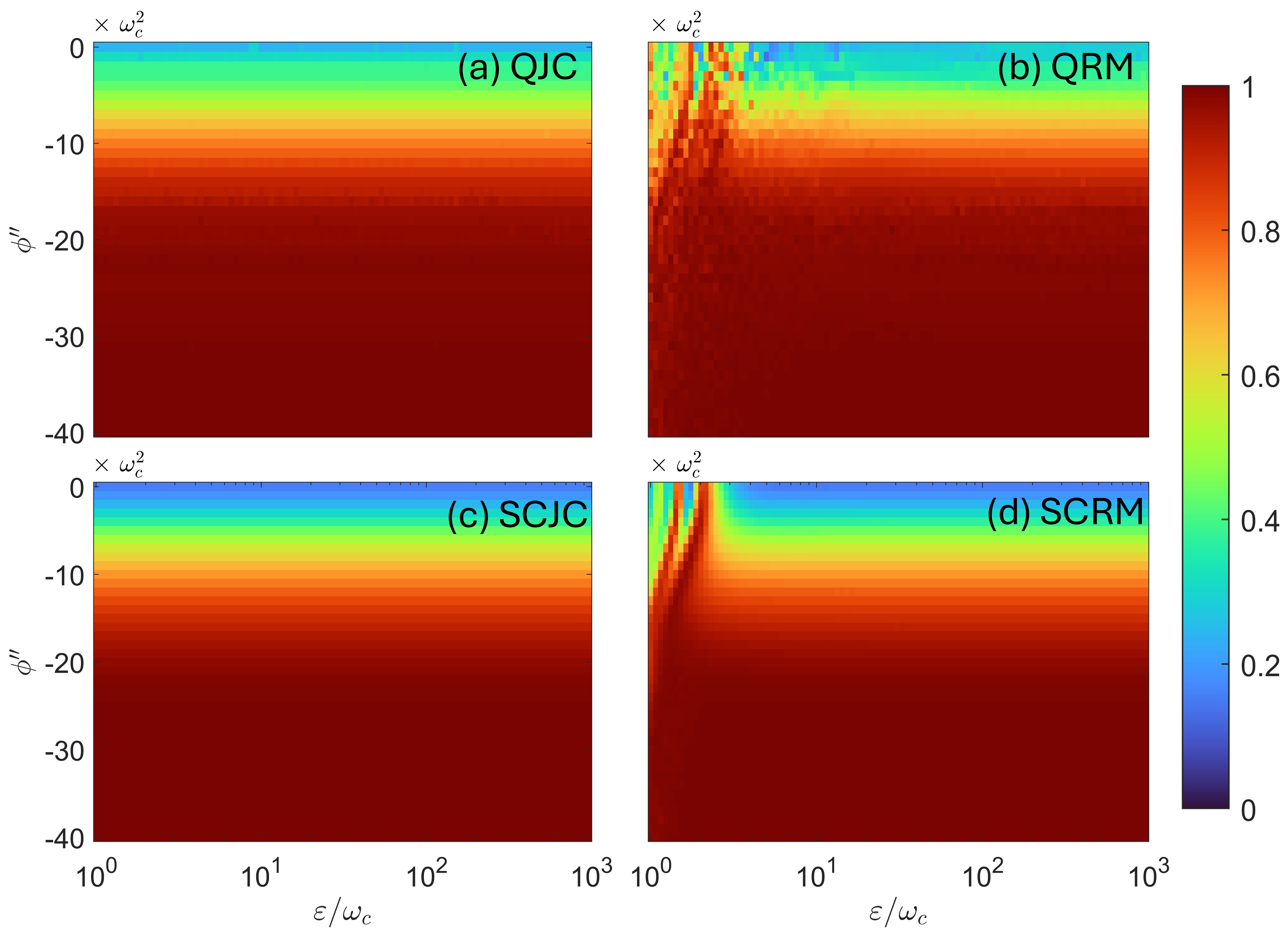}
    \caption{Two-dimensional heatmaps of the final transition probability $P_f$ as a function of the laser and energy gap resonance $\varepsilon=\omega_0$ and spectral chirp $\phi^{\prime\prime}$. The x-axis is plotted in log-scale in units of $\omega_c$. This figure is generated with the following parameters: $\Theta=10\pi$ and $A=0.22~\omega_c^{-2}$ for the (a) QJC ($|\alpha|^2=1$), (b) QRM ($|\alpha|^2=1$), (c) SCJC and (d) SCRM.}
    \label{figure3}
\end{figure}

The behavior of $P_f$, obtained from QRM, QJC, SCJC and SCRM models, as a function of the energy gap at resonant excitation $\omega_0=\varepsilon$ and the spectral chirp $\phi^{\prime\prime}$ is shown in Fig. \ref{figure3}. These heatmaps is useful for relating the dimensionless parameter
$\varepsilon/\omega_c$ to experimentally relevant hierarchy regimes.
Here, $\varepsilon$ denotes the driven two-level transition energy,
whereas $\omega_c$ represents the characteristic cutoff frequency of
the phonon or vibrational environment. The values
$\varepsilon/\omega_c \sim 1,10,10^2$, and $10^3$ should therefore be
interpreted as illustrative classes of system--bath separation rather
than exact one-to-one mappings to individual experiments.

\begin{itemize}
    \item $\varepsilon/\omega_c \sim 1$: weakly separated
    system--bath energy scales, where the driven transition is comparable to
    the dominant vibrational or phonon energy. This regime is relevant
    to vibronically resonant molecular aggregates and photosynthetic
    complexes, where electronic energy gaps can be close to selected
    intramolecular vibrational modes. Representative experimental
    probes include two-dimensional electronic spectroscopy and
    pump--probe spectroscopy of vibronically coupled molecular systems
    \cite{Tiwari2013}. Engineered low-energy solid-state platforms,
    such as silicon quantum dots with tunable valley splitting, also
    provide possible realizations of this hierarchy
    \cite{Yang2013,Hollmann2020}.

    \item $\varepsilon/\omega_c \sim 10$: moderately separated
    energy scales, corresponding to optical or electronic transitions that
    exceed local vibrational energies by roughly one order of
    magnitude. Representative examples include molecular chromophores,
    color centers, and defect-based emitters whose electronic
    transitions lie in the visible or near-infrared while the relevant
    local vibrational modes are in the tens to hundreds of meV range.
    Suggested experiments include chirped femtosecond excitation,
    transient absorption, fluorescence up-conversion, and phonon
    sideband spectroscopy of organic molecules or diamond color
    centers \cite{Davies1976,Kehayias2013}.

    \item $\varepsilon/\omega_c \sim 10^2$: strongly separated
    energy scales, characteristic of excitonic transitions in two-dimensional
    semiconductors, where eV-scale optical excitations couple to
    phonon modes in the meV to tens-of-meV range. Representative
    platforms include monolayer transition-metal dichalcogenides such
    as WSe$_2$ and MoSe$_2$, for which exciton--phonon coupling plays
    an important role in photoluminescence, linewidths, valley
    relaxation, and phonon-assisted exciton formation
    \cite{Huang2016,Chow2017}. Suitable experiments include
    chirped-pulse excitation of neutral excitons or trions,
    time-resolved photoluminescence, resonant Raman spectroscopy,
    photocurrent spectroscopy, and two-dimensional coherent
    spectroscopy.

    \item $\varepsilon/\omega_c \sim 10^3$: very strongly separated
    energy scales, naturally realized in optically driven semiconductor
    quantum dots, where the excitonic transition is in the eV range
    while the relevant acoustic-phonon cutoff is typically in the meV
    or sub-meV range. This regime is especially relevant to
    self-assembled InGaAs/GaAs or InAs/AlAs quantum dots, where
    chirped rapid adiabatic passage and phonon-assisted population
    inversion have already been demonstrated experimentally
    \cite{Simon2011,Sarkar2008,Quilter2015}. These systems provide a
    direct experimental route for testing the chirp-controlled
    inversion plateau predicted by the present open QRM and QJC models.
\end{itemize}

For trapped ions, the two-level system can be encoded in the internal electronic or hyperfine states of a single ion, while one quantized vibrational mode plays the role of the bosonic mode. A concrete realization was reported by Lv \textit{et al.}, who experimentally simulated the quantum Rabi model using a single \(^{171}{\rm Yb}^{+}\) ion. In their experiment, the hyperfine transition frequency was \(\omega_{\rm HF}=2\pi\times 12.64~{\rm GHz}\), and a radial vibrational mode with frequency \(\omega_X=2\pi\times 2.50~{\rm MHz}\) was used as the bosonic mode \cite{Lv2018QRM}. In the effective rotating-frame quantum Rabi model, the simulated parameters are controlled by the red- and blue-sideband detunings,
\begin{equation}
    \omega_1=\frac{\delta_b+\delta_r}{2},
    \qquad
    \omega_2=\frac{\delta_b-\delta_r}{2},
    \qquad
    g=\frac{\eta\Omega}{2}.
\end{equation}
In the coupling-regime experiment, Lv \textit{et al.} used \(g=2\pi\times 12.5~{\rm kHz}\) and varied the effective mode frequency over approximately \(\omega_2/2\pi \approx 20\)--\(300~{\rm kHz}\), corresponding to coupling ratios \(g/\omega_2=0.04\), \(0.6\), and \(1.2\) \cite{Lv2018QRM}. These values show that trapped ions can access the Jaynes--Cummings, ultrastrong-coupling, and deep-strong-coupling regimes relevant to the present work. Therefore, a realistic trapped-ion implementation of our chirped protocol could use effective frequencies in the tens to hundreds of kHz range, coupling strengths of order \(10~{\rm kHz}\), and chirped detuning sweeps over comparable frequency windows.

Superconducting circuit QED provides another experimentally relevant platform. In this case, the two-level system is a superconducting artificial atom coupled to a microwave resonator. Early circuit-QED proposals and experiments used microwave resonators with frequencies in the few-GHz range and demonstrated strong single-photon light--matter coupling. For example, the original circuit-QED proposal by Blais \textit{et al.} estimated a vacuum Rabi rate \(g/\pi\) of order \(100~{\rm MHz}\), corresponding to approximately \(1\%\) of the transition frequency \cite{Blais2004PRA}. Wallraff \textit{et al.} subsequently demonstrated strong coupling of a single photon to a superconducting qubit in a microwave resonator, with a resonator frequency around \(2\pi\times 6~{\rm GHz}\) and a vacuum Rabi frequency on the order of \(10~{\rm MHz}\) \cite{Wallraff2004Nature}.

We have assumed a zero-temperature phonon bath throughout this work in order to isolate the effects of field quantization, chirp, and counter-rotating light--matter interactions. At finite temperature, each phonon mode can be initiated in a P-function representation of the canonical density operator with a width parameter,
\begin{equation}
    2\sigma_k^2=\frac{1}{\exp(\hbar\omega/k_B T)-1}.
\end{equation}
Equivalently, the zero-temperature phonon bath correlation function is replaced by its finite-temperature form. Thus, finite temperature enhances the symmetric noise component of the phonon bath and is expected to increase phonon-induced dephasing and relaxation.\cite{Leggett1987RMP,Weiss2012QDS,Nazir2016QDots} For the chirped rapid adiabatic passage dynamics considered here, this should reduce the maximum achievable inversion fidelity and narrow the high fidelity chirp-controlled inversion window, particularly when \(k_B T\) becomes comparable to the relevant dressed-state splittings or phonon frequencies sampled during the pulse. In the low-temperature regime, where \(n_B(\omega,T)\ll 1\) for the dominant phonon modes, the zero-temperature approximation should remain qualitatively reliable. A quantitative finite-temperature treatment, for example using finite-temperature bath correlation functions or a thermofield-dynamics representation of the phonon bath, is left for future work.

The comparison in Fig. 3 is also informative as it clearly highlights the differences between the four models clearly which is important to note when considering which model to use. By comparing the two quantum models in Fig. \ref{figure3}(a) and (b), the parameter regime where RWA is valid seems to appear after $\frac{\varepsilon}{\omega_c}\gtrsim50$. Interestingly, a general feature across all four models is that decreasing $\phi^{\prime\prime}$ from $0~\text{$\omega_c$}^2$ to $-20~\text{$\omega_c$}^2$ can steadily increase the $P_f$ and as $\phi^{\prime\prime}<-20~\text{$\omega_c$}^2$, $P_f\geq 0.9$. This suggests that the ideal sweep rate for adiabaticity is already achieved with $\phi^{\prime\prime}=-20~\text{$\omega_c$}^2$. The $P_f$ for QJC and SCJC are  independent of the energy gap at resonant excitation as expected due to the transformation into the rotating frame that only takes into account the detuning. However, accounting for the counter-rotating terms, the QRM and SCRM models show that the system can achieve high $P_f$ with low magnitude spectral chirp $|\phi^{\prime\prime}|$ when $\frac{\varepsilon}{\omega_c}\lesssim40$. It is important to note that even if $P_f$ is similar for all four models, the time-dependent population $P_1(t)$ could vary drastically, especially around $\omega_ct=0$. This comparison across all four models identifies parameter regimes in which commonly used semiclassical or RWA descriptions give reliable final populations but incorrect transient dynamics, a distinction that is directly relevant to interpreting chirped-pulse experiments in noisy solid-state emitters and quantum simulators.

In summary, we have benchmarked chirped rapid adiabatic passage in open Jaynes--Cummings and Rabi models against their semiclassical counterparts, with and without the rotating-wave approximation, using a variational multiple-Davydov $D_2$ treatment of an initially coherent field with very large as well as low mean photon number. The semiclassical description remains reliable in the large-$|\alpha|^2$, small-$\Theta$ regime but breaks down at low photon number, where field quantization and fluctuations become essential. Within the present models, the QRM retains both field quantization and counter-rotating processes and therefore provides the most complete of the four descriptions considered. The QRM also reveals a robust chirp-controlled inversion window: for sufficiently negative spectral chirp, near-unity transfer becomes largely insensitive to pulse area and spin--phonon coupling (e.g., $\phi''=-40~\omega_c^2$ gives $P_f\simeq 1$ once $\Theta\gtrsim 3\pi$), while $\phi''\lesssim -20~\omega_c^2$ already yields $P_f\gtrsim 0.9$ over a broad range of transition energies.
In contrast, RWA-based JC descriptions incorrectly depict indifference of $P_f$ with respect to resonant excitation and can fail when computing the time evolution of $P_1(t)$ at large pulse areas where counter-rotating channels contribute significantly.
These results elucidate when semiclassical and RWA approximations are trustworthy and identify experimentally favorable chirp protocols for high-fidelity population transfer in noisy solid-state emitters. 

Beyond accuracy, the key experimental implication is a chirp-selected \emph{inversion plateau} that relaxes calibration requirements. High-fidelity preparation persists despite substantial variations in pulse area, detuning, and noise strength. At the same time, sweeping $(\phi'',\Theta, |\alpha|^2)$ provides a sharp, platform-independent benchmark for the onset of quantization and counter-rotating-induced deviations, preventing systematic misinterpretation when fitting data with semiclassical or RWA models.
Our predictions are directly testable in trapped-ion quantum simulators \cite{Pedernales,Dingshun} and superconducting circuit QED \cite{Braum,Felicetti}, where chirped driving field and in situ tunability of detuning and light-matter coupling enable a direct validation of the chirp-protected inversion window in a noisy environment and the regimes where semiclassical or RWA descriptions fail. A multimode photon field with a Gaussian spectrum would be the next step for future work.

\section*{ACKNOWLEDGEMENTS} Support from the Singapore Ministry of Education Academic Research Fund Tier 1 (Grant No. RG2/24) is gratefully acknowledged. M.F.G acknowledges support from the National Natural Science Foundation of China (No. 22373028).

\section*{ASSOCIATED CONTENT}
\subsection*{Supporting Information}
Derivation of the displaced effective single-mode quantum Hamiltonian and its semiclassical large-\(|\alpha|^2\) limit; coherent-state phase convention. Equations of motion derived from the Dirac-Frenkel time-dependent variational principle for the variational parameters of the mD$_2$ wavefunction and the convergence of numerical accuracy is included. The effective single-mode representation of a multimode chirped Gaussian pulse can be found in the Supporting Information as well.

\newpage
\section*{AUTHOR INFORMATION} 
\noindent \textbf{Justin Zhengjie Tan} --- \textit{School of Materials Science and Engineering, Nanyang Technological University, Singapore 639798, Singapore}; \url{orcid.org/0009-0004-7193-9880} \par\smallskip \noindent \textbf{Frank Gro{\ss}mann} --- \textit{Institute for Theoretical Physics, Technische Universit\"at Dresden, 01062 Dresden, Germany}; \url{orcid.org/0000-0003-1415-8426} \par\smallskip \noindent \textbf{Yiying Yan} --- \textit{School of Materials Science and Engineering, Nanyang Technological University, Singapore 639798, Singapore; Department of Physics, School of Science, Zhejiang University of Science and Technology, Hangzhou 310023, China}; \url{orcid.org/0000-0003-4396-7265} \par\smallskip \noindent \textbf{Maxim Gelin} --- \textit{School of Science, Hangzhou Dianzi University, Hangzhou 310018, China}; \url{orcid.org/0000-0003-3092-3343} \par\smallskip \noindent \textbf{Yang Zhao} --- \textit{School of Materials Science and Engineering, Nanyang Technological University, Singapore 639798, Singapore}; \url{orcid.org/0000-0002-7916-8687}; Email: \url{YZhao@ntu.edu.sg}

\end{document}